\documentclass[preprint,3p,twocolumn,11pt]{elsarticle}

\usepackage{amssymb}
\usepackage{amsmath}
\usepackage{lineno}
\usepackage[labelfont=bf]{caption}

\usepackage{float}
\usepackage{graphicx}
\usepackage{siunitx}
\usepackage{color,soul}
\usepackage[dvipsnames]{xcolor}
\usepackage[version=4]{mhchem}

\usepackage[
   colorlinks = true,
   linkcolor = blue,
   citecolor = blue,
   filecolor = blue,
   urlcolor = blue
]{hyperref}

\usepackage{cleveref}

\usepackage{natbib}
\biboptions{sort&compress,longnamesfirst}

\newcommand{\bnmr}{$\beta$-NMR}

\newcommand{\msr}{$\mu$SR}
\newcommand{\slr}{SLR}
\newcommand{\efg}{EFG}

\newcommand{\eli}{\textsuperscript{8}{Li}}
\newcommand{\elip}{\textsuperscript{8}{Li}\textsuperscript{+}}

\newcommand{\lip}{{Li}\textsuperscript{+}}

\newcommand{\tcrs}{1\textit{T}-CrSe$_2$}
\newcommand{\crs}{CrSe$_2$}

\journal{arXiv.org}

\begin{document}

\begin{frontmatter}

\title{Investigation of Ionic and Anomalous Magnetic Behavior in CrSe$_2$ Using $^8$Li $\beta$-NMR}

\author[label1,label2]{John O. Ticknor}
\author[label3]{Izumi Umegaki}
\author[label1,label2]{Ryan M. L. McFadden}
\author[label2,label4]{Aris Chatzichristos}
\author[label2,label4]{Derek Fujimoto}
\author[label1,label2]{Victoria L. Karner}
\author[label2,label4,label5]{Robert F. Kiefl}
\author[label6]{Shintaro Kobayashi}
\author[label5]{C. D. Philip Levy}
\author[label5]{Ruohong Li}
\author[label5]{Gerald D. Morris}
\author[label5]{Matthew R. Pearson}
\author[label7,label8]{Kazuyoshi Yoshimura}
\author[label9,label10,label11,label3]{Jun Sugiyama}
\author[label1,label2,label5]{W. Andrew MacFarlane$^{*,}$}

\address[label1]{Department of Chemistry, University of British Columbia, Vancouver, BC V6T 1Z1, Canada}
\address[label2]{Stewart Blusson Quantum Matter Institute, University of British Columbia, Vancouver, BC V6T 1Z4, Canada}
\address[label3]{Toyota Central Research and Developmental Laboratories Inc., Nagakute, Aichi 480-1192, Japan}
\address[label4]{Department of Physics and Astronomy, University of British Columbia, Vancouver, BC V6T 1Z1, Canada}
\address[label5]{TRIUMF, Vancouver, BC V6T 2A3, Canada}
\address[label6]{Japan Synchrotron Radiation Institute, SPring-8, Kouto, Hyogo 679-5198, Japan}
\address[label7]{Department of Chemistry, Graduate School of Science, Kyoto 
University, Kyoto 606-8502, Japan}
\address[label8]{Research Center for Low Temperature and Material Sciences, Kyoto 
University, Kyoto 606-8501, Japan}
\address[label9]{CROSS Neutron Science and Technology Center, Tokai, Ibaraki 319-1106, Japan}
\address[label10]{Advanced Science Research Center, Japan Atomic Energy Agency, Tokai, Ibaraki 319-1195, Japan}
\address[label11]{Institute of Materials Structure Science, High Energy Accelerator Research Organization, Tsukuba, Ibaraki 305-0801, Japan}

\tnotetext[t1]{Corresponding Author: wam@chem.ubc.ca}

\begin{abstract}

\noindent We have studied a mosaic of \tcrs\ single crystals using $\beta$-detected nuclear magnetic resonance of \eli\ from 4 to 300 K. We identify two broad resonances that show no evidence of quadrupolar splitting, indicating two magnetically distinct environments for the implanted ion. We observe stretched exponential spin lattice relaxation and a corresponding rate ($1/T_1$) that increases monotonically above 200 K, consistent with the onset of ionic diffusion. A pronounced maximum in $1/T_1$ is observed at the low temperature magnetic transition near 20 K. Between these limits, $1/T_1$ instead exhibits a broad minimum with an anomalous absence of strong features in the vicinity of structural and magnetic transitions between 150 and 200 K. Together, the results suggest \elip\ site occupation within the van der Waals gap between \crs\ trilayers. Possible origins of the two environments are discussed.

\end{abstract}

\end{frontmatter}

\section*{Introduction}

\begin{figure}[H]
 \centering
 \includegraphics[width=\columnwidth]{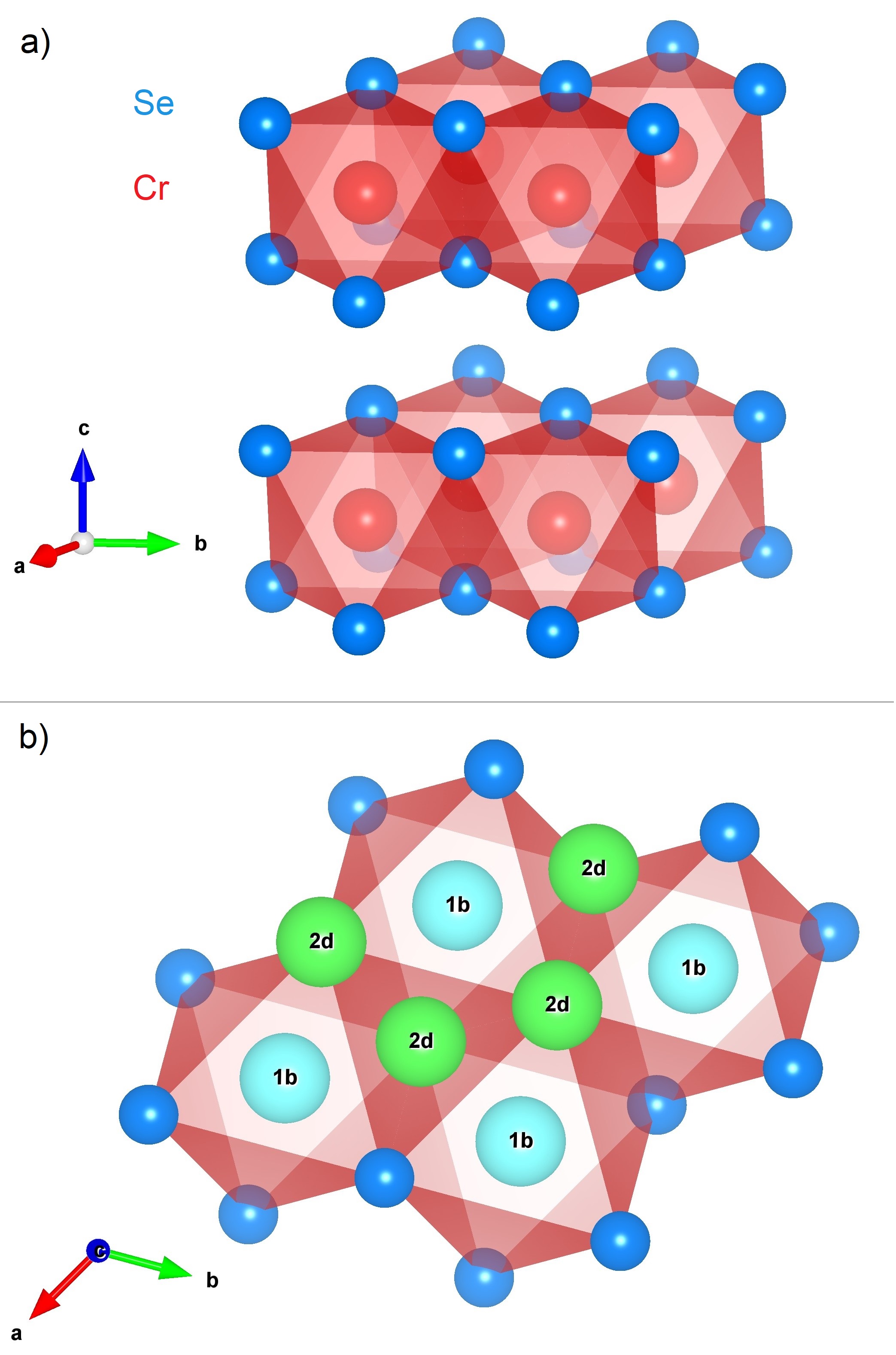}
 \caption{The \crs\ structure (space group \textit{P}$\overline{3}$\textit{m}1) visualized using VESTA \cite{2011-Momma}. (a) The stacking of layers of edge-sharing CrSe$_6$ octahedra gives an overall trigonal structure. Intercalation of small guest ions is permitted by space between the CrSe$_2$ layers. (b) The high symmetry interstitial sites in the vdW gap: the quasioctahedral ($1b$) and the quasitetrahedral ($2d$). The $1b$ site at (0, 0, 1/2) is the Li location in the fully intercalated LiCrSe$_2$ \cite{2018-Sugiyama}.}
 \label{fig:CrSe2}
\end{figure}

Transition metal dichalcogenides (TMDs) are a well known class of two-dimensional (2D) materials composed of weakly van der Waals bound stacks of strongly coordinated triangular lattice layers \cite{1976-Levy,1987-Friend}. They are good intercalation hosts for guest species such as alkali cations, enabling (for example) applications in solid state lithium ion batteries \cite{1978-Whittingham,2016-Cook,2014-Pumera}. In contrast to TMDs such as MoS$_2$, \crs\ is not well known and poorly understood. To date, its only practical synthesis is based on de-intercalating a stable parent compound (e.g., KCrSe$_2$) \cite{1980-Bruggen}. The crystal structure of \tcrs\ is illustrated in \Cref{fig:CrSe2}a, showing triatomic layers of edge-sharing CrSe$_6$ octahedra between interlayer van der Waals (vdW) gaps with vacant alkali sites shown in \Cref{fig:CrSe2}b.

%% FIGURE %%

%% FIGURE %%

Resistivity measurements indicate that \crs\ remains metallic down to $\sim 2$ K \cite{2014-Kobayashi}. TMD metals are generally susceptible to characteristic 2D electronic instabilities, such as charge density waves (CDW). For example, closely related TiSe$_2$ and VSe$_2$ have CDW transitions around 200 and 100 K, respectively \cite{2016-Sugiyama}. In contrast, \crs\ does not appear to exhibit a CDW, but the partially filled Cr $d$ shell adds the possibility of local moment magnetism. The electronic structure of the nominal Cr$^{4+}$ valence is both unusual and unstable. Due to a negative charge transfer energy, it may disproportionate into Cr$^{3+}$ and ligand Se$^{2-}$ holes. Its threefold $t_{2g}$ orbital degeneracy also makes it unstable to orbital ordering \cite{2014-Kobayashi}. Recent muon spin rotation (\msr) measurements suggest incommensurate spin density wave (SDW) order below $T_{N1} = 157$ K \cite{2016-Sugiyama}. Furthermore, this state was found to be highly sensitive to S/Se substitution \cite{2019-Kobayashi}. While the muon is a good probe of magnetic behavior, investigation of the dilute-limit \lip\ dynamics can also provide important details of the short-range, microscopic \lip\ motion \cite{2017-McFadden}. This prompts the use of a similar, but complementary probe: the short-lived radioisotope \eli.

Nuclear magnetic resonance (NMR) offers a powerful local probe of the electronic and magnetic properties of TMDs \cite{1992-Naito}, as well as the intercalant dynamics \cite{1995-Prigge,2006-Wilkening,2009-Bensch}. Here, we use a variant technique (\bnmr), where the signal is detected from the anisotropic \eli\ radioactive $\beta$-decay \cite{2015-MacFarlane}. As with conventional NMR \cite{2004-Grey}, wide lines and fast spin lattice relaxation (SLR) make \bnmr\ for strongly magnetic materials such as \crs\ very challenging. We investigated single crystals of \crs\ using implanted \elip\ ions in the dilute limit, spanning a temperature range of 4 to 300 K. We find two broad NMR lines, reflecting two distinct magnetic environments. One line disappears between 150 and 200 K, the range where CrSe$_2$ exhibits structural transitions. Above 200 K, the steady increase of $1/T_1$ is suggestive of \elip\ diffusion.

\section*{Experimental}

\begin{figure}[H]
\centering
  \includegraphics[width=\columnwidth]{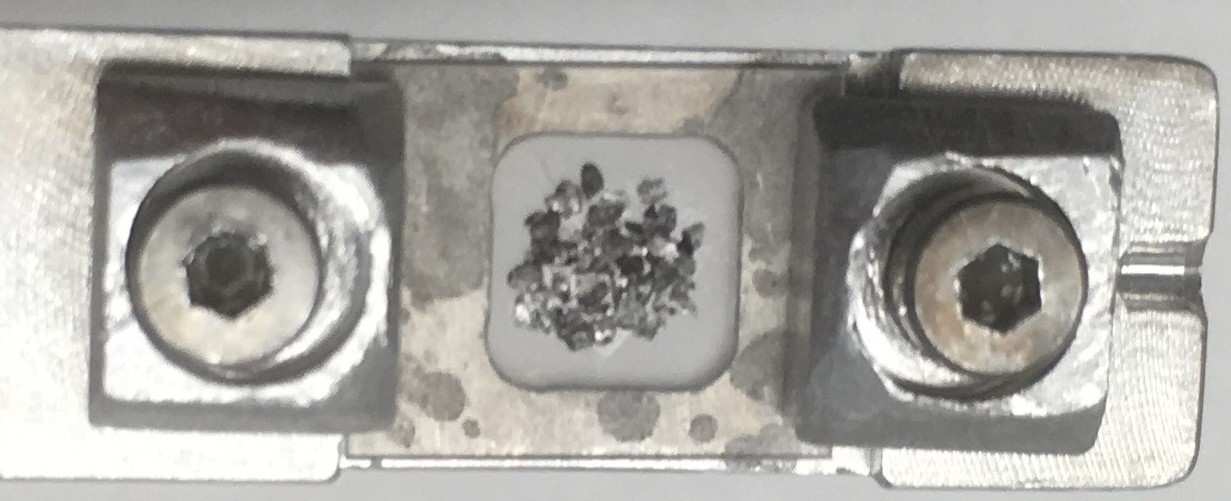}
  \caption{A photograph of the mosaic of \crs\ single crystals affixed to a sapphire plate (Crystal GmbH, Berlin) using Apiezon-L grease (M \& I Materials, Manchester). The tight packing and highly focused \elip\ beam spot used in the experiment minimizes the \bnmr\ signal from the surrounding materials.}
  \label{fig:Sample}
\end{figure}

\crs\ single crystals were prepared by de-intercalation of an alkali precursor, K$_x$CrSe$_2$. As previously reported \cite{2014-Kobayashi}, Bragg-Brentano X-ray diffraction, transmission electron microscopy, and energy-dispersive X-ray spectroscopy measurements confirmed phase purity with near-perfect CrSe$_2$ stoichiometry and negligible remaining alkali. A tightly packed mosaic of $\sim$ 50 small ($\sim$ 0.3 mm $\times$ 0.3 mm $\times$ 0.1 mm) crystals was affixed to a sapphire plate with a small amount of ultra-high vacuum (UHV) compatible grease (Apiezon-L; M \& I Materials, Manchester, UK), see \Cref{fig:Sample}. The sample was oriented so that the beam was implanted along the crystalline $c$-axis.

%% FIGURE %%

%% FIGURE %%

$\beta$-NMR measurements made use of the Isotope Separator and Accelerator (ISAC) facility at TRIUMF in Vancouver, Canada \cite{2014-Dilling}. Here, a 500 MeV proton beam supplies a continuous source of radioisotopes from nuclear reactions in a tantalum production target. Surface ionization, followed by high resolution mass separation, provides an intense, isotopically pure \elip\ ($\sim 10^{7}$ s$^{-1}$) ion beam, here with an energy of 19 keV. The beam is then spin polarized in-flight in a three step process: neutralization via charge exchange in a rubidium vapor cell, optical pumping of the neutral atom beam with resonant circularly polarized laser light ($D$ line) to polarize the electronic and nuclear spin, and ionization in a helium gas cell \cite{2010-Levy}. The re-ionized spin-polarized beam is then delivered to the spectrometer, which is based on a UHV ($< 10^{-9}$~Torr), 9 Tesla superconducting solenoid (Oxford Instruments). The spin polarization of the delivered ions is parallel to both the beam direction \textit{and} the static applied field $B_0$, defining the laboratory frame's $z$-axis. At this implantation energy, the mean implantation depth in \crs\ is approximately 94 nm according to SRIM Monte Carlo simulations \cite{2010-Ziegler}.

Similar to \msr, the spin polarization is detected by the radioactive $\beta$-decay. Here, the experimental $\beta$-decay asymmetry, $A(t)$, is determined from the time-resolved $\beta$ rates ($N_{F}$ and $N_{B}$) in two detectors on opposite sides of the sample, the forward $F$ counter (downstream of the sample) and the annular backward (upstream) $B$ counter, by:

%% EQUATION %%

\begin{equation}
\label{eq:asym}  
A(t) = \frac{N_{F}(t) - N_{B}(t)}{N_{F}(t) + N_{B}(t)} = A_{0}p_{z}(t) = A_{0} \left( \frac{\langle I_{z}(t) \rangle}{I} \right)
\end{equation}

%% EQUATION %%

The nuclear spin polarization $p_{z}(t)$ is the average of the projection of the ($I=2$) \eli\ nuclear spin along $z$, $\langle I_{z}(t) \rangle$, over the ensemble of all the implanted ions. The proportionality constant, $A_0 \approx 0.1$, is determined by the properties of the $\beta$-decay, the  detector geometry, and other experiment-specific details. To a good approximation, $A_0$ is independent of temperature. 

The \eli\ lifetime ($\tau = 1.21$ s) is much longer than the muon, making \bnmr\ sensitive to behavior occurring on a separate timescale compared to \msr. Also, the detection scheme of \bnmr\ makes it far more sensitive than conventional NMR; significantly fewer nuclei are required for sufficient signal, and one can measure samples that would be impossible to study with ordinary NMR (e.g., thin films) \cite{2015-MacFarlane}. In the present case, we are interested in the mobility of isolated interstitial \lip\ in \crs.

The main observables in \bnmr\ are the resonance spectrum and the spin lattice relaxation (SLR). This is similar to conventional NMR, but they are measured in a very different way. In conventional NMR, a sequence of RF pulses disturbs the nuclear magnetic sublevel populations and their subsequent recovery to thermal equilibrium is monitored to obtain the SLR rate $1/T_1$. In \bnmr, the probes are implanted in a spin state initially far from equilibrium, and one can simply observe the relaxation. We implant a 1 second pulse of spin-polarized \elip\ (i.e., the beam-on window in \Cref{fig:SLR_Res}a), followed by a beam-off period of about 10 \eli\ lifetimes. In \Cref{fig:SLR_Res}a, the relaxation is so fast that only the first 3 seconds of the beam-off period are shown. During the entire cycle, the asymmetry $A(t)$ is proportional to the total spin polarization of all \eli\ in the sample. During the pulse, newly implanted \eli\ continuously replenishes the decaying polarization, and the recovery is to a dynamic equilibrium value. After the pulse, the polarization simply relaxes towards the thermal equilibrium value near zero. This leads to the pronounced kink in the \bnmr\ recovery curves shown in \Cref{fig:SLR_Res}a. This cycle is repeated for alternating laser helicity (the sense of \eli\ polarization), giving a total measurement time of about 20 minutes. The two helicities are then combined to form the averaged asymmetry which is then fit to a relaxation model as detailed below. No RF pulses are used in the \slr\ measurement, so it has no spectral resolution; it captures the relaxation of \textit{all} the \eli\ in the sample (i.e., the full spectrum), even for nuclei with resonances too broad to observe.

The \bnmr\ resonance spectrum  (see \Cref{fig:SLR_Res}b) is measured using a continuous beam of \eli, that yields the time average dynamic equilibrium polarization. Opposite to conventional NMR, a continuous wave (CW) RF excitation is used rather than RF pulses. At a specific frequency, the asymmetry is averaged for an integration time, then stepped to the next value. Here, we use a range 41.08-41.42 MHz (encompassing the \eli\ Larmor frequency) with a 2 kHz step and a 1 s integration time. When the RF excitation matches the NMR frequency, the \eli\ spins precess rapidly about it, and the asymmetry is averaged out. Resonances are thus observed as a reduction of the asymmetry. To obtain better statistics, the frequency scan is repeated a number of times, alternating both helicity and scan direction to reduce systematics, for a total acquisition time of about 1 hour.

Although the RF is monochromatic, the resulting spectrum is power broadened: it is convoluted with a Lorentzian of width $\gamma H_1$, where $H_1$ is the RF field amplitude. In practice, $H_1$ is selected as a compromise between amplitude (larger $H_1$) and resolution (smaller $H_1$). Here, we use $H_1$ near the maximum value of 0.5 G, because the resonances are so wide that power broadening is negligible compared to the intrinsic width. Note that the spectrum is also \textit{averaged} over a rather long integration time, in contrast to the high speed ``snapshot'' static spectra obtained in pulsed NMR.

While the advantages of pulsed RF in NMR are widely known, they have not been applied extensively in \bnmr. Effective pulses, particularly for broad lines, require much higher $H_1$. Besides this strictly technical barrier, the measurement time is limited statistically by the radioactive lifetime $\tau$: one needs enough time for the \eli\ to decay in order to observe the asymmetry.

Because the polarization is produced externally by optical pumping and measured by the $\beta$-decay, the signal-to-noise ratio is independent of the resonance frequency (applied static field, $H_0$) \cite{2014-Morris}, as opposed to conventional NMR, where it grows as $H_0^2$. As a result, the field may be freely varied. Here, for example, we determine its effect on $1/T_1$. However, both the incident ion and $\beta$ electron trajectories are sensitive to the field. The former necessitates a time consuming process of fine-tuning the beam transport at each field, while the latter reduces the $\beta$ detector efficiency below 1 Tesla.

\section*{Results}

\begin{figure}[H]
 \centering
 \includegraphics[width=\columnwidth]{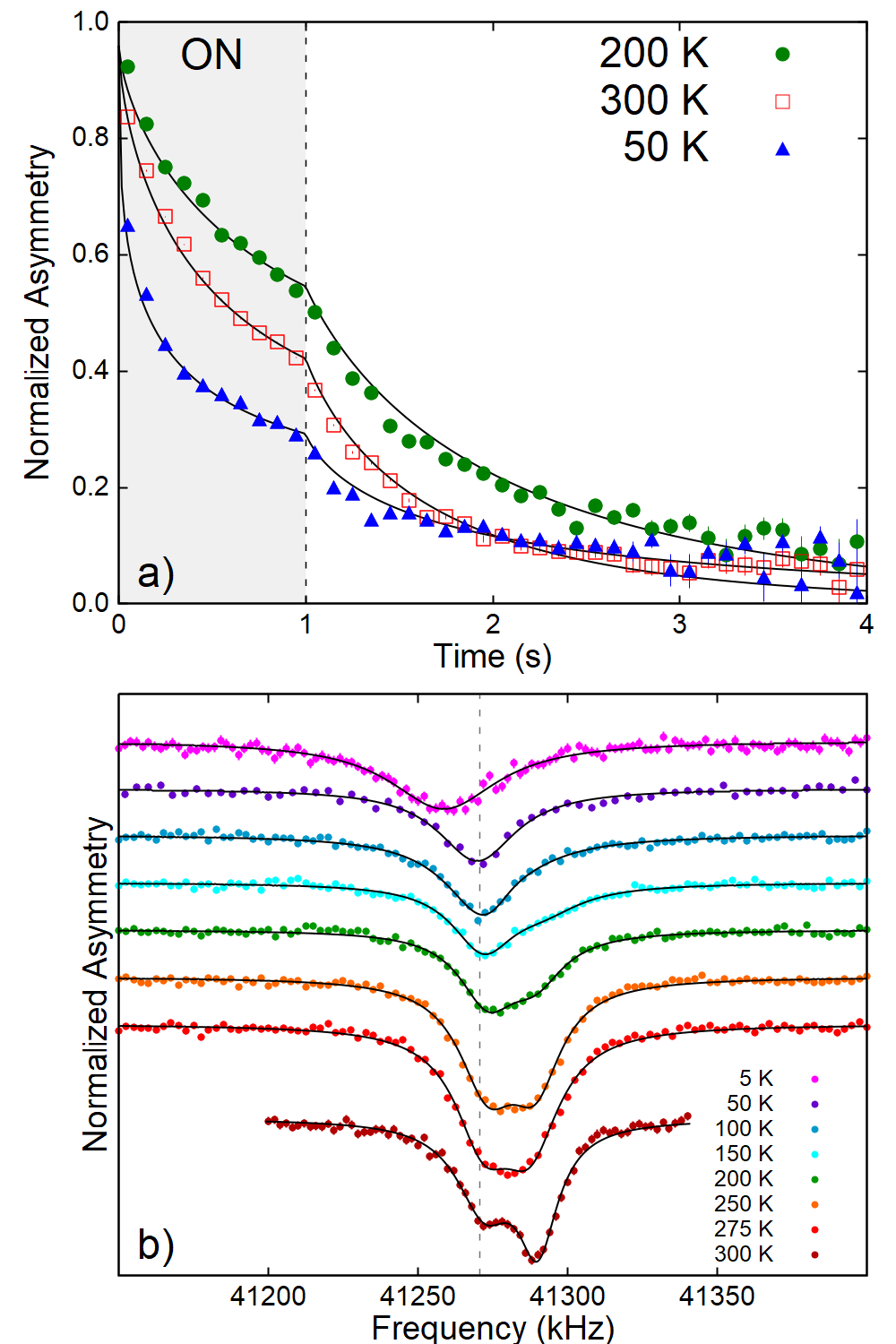}
 \caption{a) $^8$Li$^+$ \slr\ data in \crs\ at selected temperatures with $B_0 = 6.55$ T. The one second beam pulse is indicated by the grey-shaded (ON) area. The solid lines depict best fits to a single-component stretched exponential. The data are binned by a factor of 10 for clarity. An overall non-monotononic temperature dependence of the \slr\ rate is clear. b) Time-averaged continuous wave (CW) resonances at various temperatures. The vertical grey dashed line indicates the MgO reference. At 300 K, the clearly distinguished broad resonance lines indicate two magnetically inequivalent environments. All spectra are offset vertically for clarity and normalized to their baseline asymmetry \cite{2009-Hossain}.
 } 
 \label{fig:SLR_Res}
\end{figure}

Representative time-resolved \slr\ data are shown in \Cref{fig:SLR_Res}a. Following the beam pulse, $A(t)$ relaxes to its near-zero equilibrium value. The rate of this decay is the \slr\ rate, $1/T_1$. An important feature of the data is that the statistical uncertainty is minimal near the trailing edge of the beam pulse; while after the pulse, it increases exponentially with the \eli\ lifetime \cite{2015-MacFarlane}.  

%% FIGURE %%

%% FIGURE %%

From the raw data, a non-monotonic temperature dependence of the relaxation rate is immediately seen. At all measured temperatures and fields, the \slr\ spectra are not well-described by single exponential relaxation. Thus, we adopt the phenomenological stretched exponential model. Specifically, for an individual \eli\ arriving at time $t'$, the spin polarization for $t>t'$ is given by $p_z(t,t') = p_0 \exp \left \{ \left [ (t-t')/(T_1) \right ]^{\beta} \right \}$, where $\beta$ is the stretching exponent. This $p_z(t,t')$ can be considered as a weighted average of exponentials with a distribution of relaxation rates that broadens rapidly as $\beta$ decreases below the single exponential limit, $\beta=1$ \cite{1980-Lindsey}.

%% FIGURE %%

\begin{figure}[H]
 \centering
 \includegraphics[width=\columnwidth]{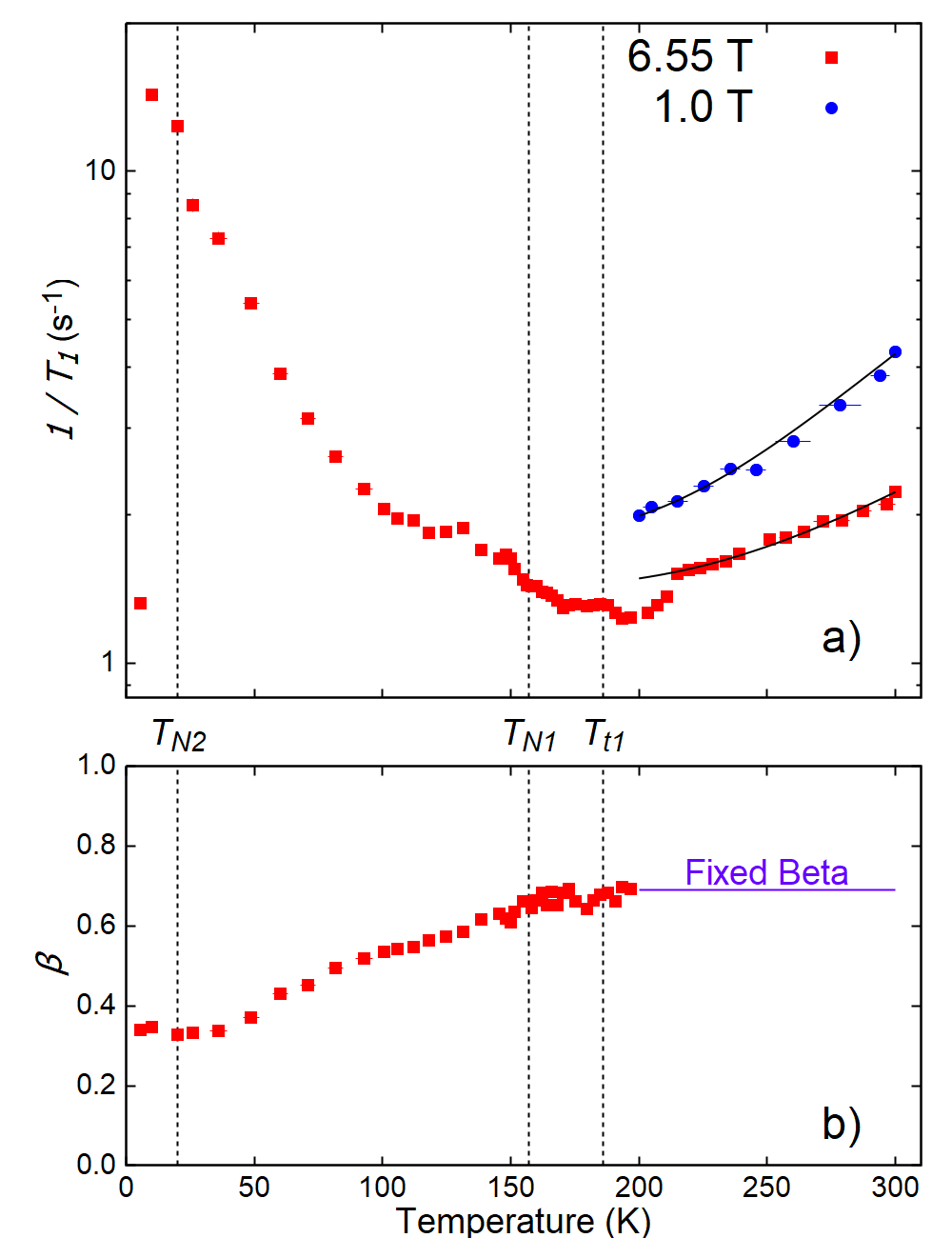}
 \caption{a) \slr\ rates $1/T_1$ and b) stretching exponent $\beta$ as a function of temperature from stretched exponential fits to \eli\ \slr\ at 6.55 and 1.0 T in \crs. Note that $\beta$ is fixed for both applied fields ($H_0$). Antiferromagnetic phase transitions are indicated by $T_{N1}$ and $T_{N2}$ \cite{2016-Sugiyama}. Structural transitions are labeled $T_{t1}$ and $T_{N1} = T_{t2}$ \cite{2014-Kobayashi}. The \elip\ sensitivity to spin fluctuations near the low temperature antiferromagnetic transition ($T_{N2}$) is evidenced by an apparent maximum in the \slr\ rate, $1/T_1$. In contrast, a broad minimum in $1/T_1$ is observed near $T_{N1}$ and $T_{t1}$. Above 200 K, $1/T_1$ grows monotonically with temperature, as expected for diffusion of \elip. The curves are fits to \Cref{eq:1/T1arr}, as discussed in the text.}
 \label{fig:T1_Beta}
\end{figure}

%% FIGURE %%

\newpage

%% FIGURE %%

\begin{figure}[H]
 \centering
 \includegraphics[width=\columnwidth]{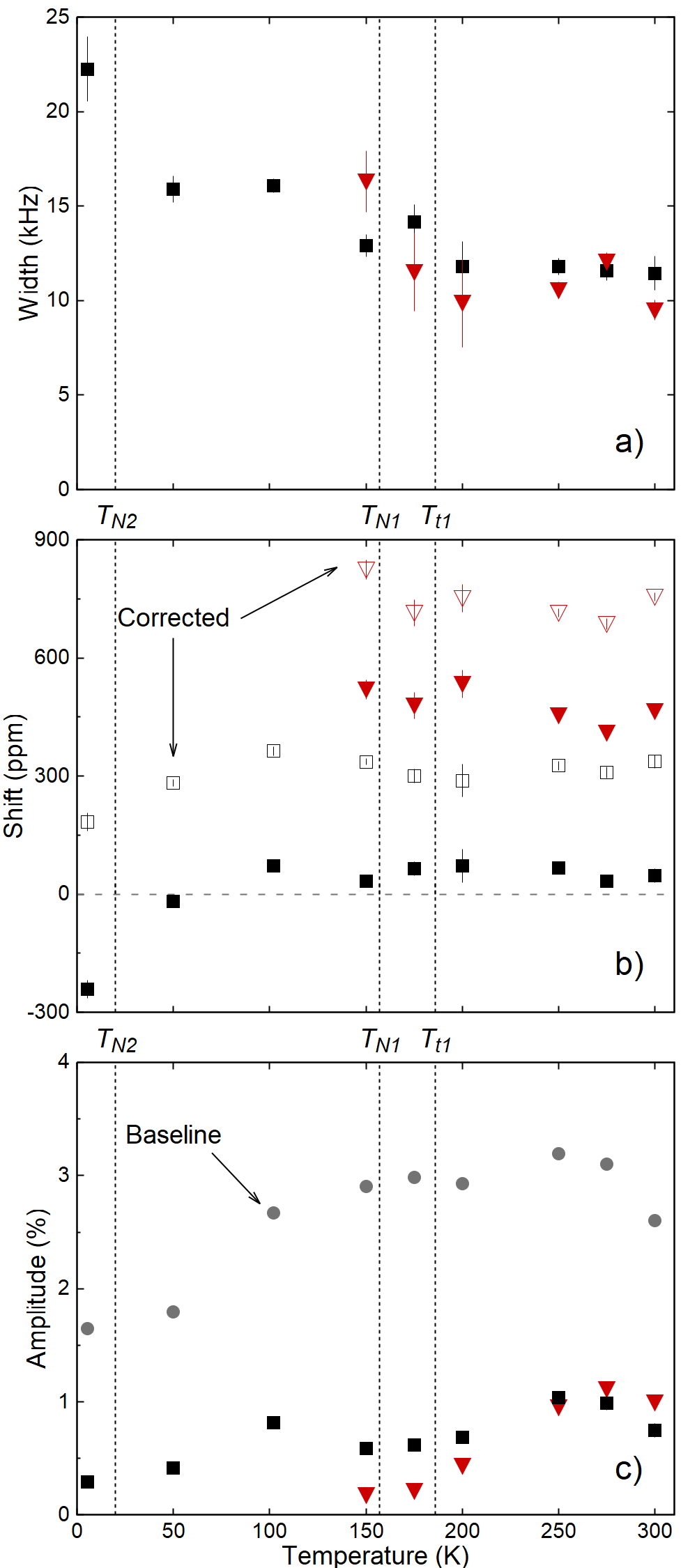}
 \caption{a) Resonance linewidths (FWHM), b) raw and demagnetization corrected shifts, and c) resonance amplitudes and off-resonance baseline asymmetries (grey-filled circles) from Lorentzian fits to the spectra in \Cref{fig:SLR_Res}b. The two resonance lines are distinguished by black squares and red triangles. Above 50 K, the demagnetization correction is approximately $+300$ ppm; independent of $T$. Deviation from this at the lowest temperature is due to the rise in susceptibility.\cite{2014-Kobayashi} The grey-colored dashed line in (b) indicates zero shift from MgO. All panels indicate aforementioned antiferromagnetic ($T_{N1}$,$T_{N2}$) and structural ($T_{t1}$,$T_{t2} = T_{N1}$) phase transitions.
 }
 \label{fig:Width_Shift_Amp}
\end{figure}

%% FIGURE %%

The data both during and after the pulse is fit by convoluting $p_z(t,t')$ with the square beam pulse \cite{2006-Salman,2015-MacFarlane-2}. This generates the curved lines in \Cref{fig:SLR_Res}a.

The temperature dependence of the fit parameters ($1/T_1$ and $\beta$) is shown in \Cref{fig:T1_Beta}. Fits for all temperatures and fields shared a fixed initial asymmetry, $A_0$ = 0.0973, while $\beta$ and $1 / T_1$ were allowed to vary, except above 200 K, where $\beta$ was approximately constant. In this range, $\beta$ was fixed to its average value 0.69 to avoid overparameterization. Overall fit quality ($\chi ^2$ = 1.01) was good. As shown in \Cref{fig:T1_Beta}a, $1/T_1$ goes through a low temperature peak around 20 K, followed by a broad minimum in the range 150 to 200 K and a subsequent high temperature increase. A narrow temperature scan above 200 K was also carried out in a lower applied field, $B_0 = 1$ T. Here, the \slr\ rates are significantly faster than at 6.55 T and follow the same trend. The resonance spectra in \crs\ (\Cref{fig:SLR_Res}b) show a single, broad resonance at low temperature. Two distinct lines appear on warming to 300 K.

The spectra were fitted with a sum of two Lorentzians above 150 K, and a single Lorentzian below. The linewidths are narrowest at the highest temperature, but they are still very broad with a FWHM (full width at half maximum) $\sim 20$ kHz. The low-temperature line persists on warming, and its shift relative to MgO is small, except below 50 K where it becomes negative. The second resonance, evident above 150 K, is strongly positively shifted. It gradually broadens and diminishes in amplitude with reduced temperature. We calculate the relative shift in parts per million (ppm) as $\delta = 10^6 (\nu - \nu_{\mathrm{ref}})/\nu_{\mathrm{ref}}$. The reference frequency $\nu_{\mathrm{ref}}$ was determined by calibration measurements in a single crystal of MgO \cite{2014-MacFarlane}. The shift is a measure of the internal field at the \eli\ nucleus in the sample. To isolate the contribution of the hyperfine field, we must account for demagnetization. In the linear response regime of the paramagnetic state, the demagnetization field is proportional to the applied field, and the corrected shift is
$\delta_c = \delta + \delta_{\mathrm{demag}}$, where

%% EQUATION %%

\begin{equation}
\label{eq:demag-corr}
\delta_{\mathrm{demag}} = 4 \pi \left( N - \frac{1}{3} \right) \chi_{0}(T)
\end{equation}

%% EQUATION %%

\begin{figure}[H]
    \centering
    \includegraphics[width=\columnwidth]{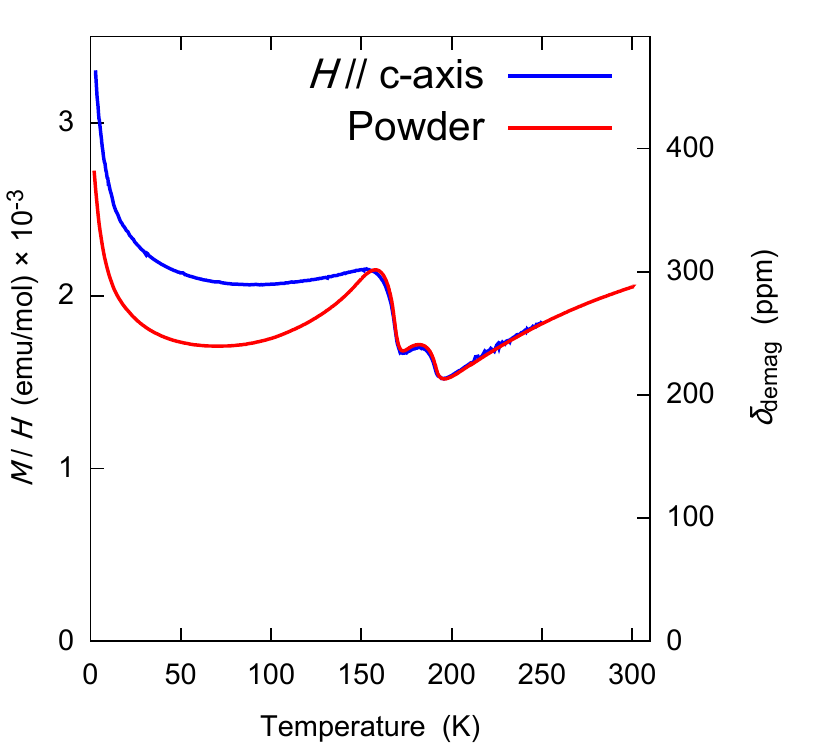}
    \caption{The (1 Tesla) field-cooled molar magnetization ($M/H$) of randomly oriented powder and single crystal \crs. This was measured and previously reported by Kobayashi, et.~al.~\cite{2014-Kobayashi} Note that above the magnetic transition at 157 K, this is the susceptibility $\chi_0$, but below this the response is somewhat nonlinear and hysteric. The right-hand axis shows the demagnetization correction $\delta_{\mathrm{demag}}$, defined in \Cref{eq:demag-corr}, which was used to calculate the corrected shifts.}
    \label{fig:Suscept}
\end{figure}

Here, $\chi_{0}(T)$ represents the dimensionless (volume) susceptibility (CGS units), and $N$ is the shape-dependent demagnetizing factor. For $B_0$ perpendicular to one of the platelet crystals, we estimate $N \approx 0.745$ assuming ellipsoidal crystallites \cite{1945-Osborn}. As the crystals are not perfectly ellipsoidal, the demagnetization field will be somewhat inhomogeneous, yielding a source of broadening. With this and $\chi_{0}$ measured at 1.0 T (\Cref{fig:Suscept}), we estimate the demagnetization correction is about +300 ppm. The raw and corrected shifts are shown in \Cref{fig:Width_Shift_Amp}b.

%% FIGURE %%

%% FIGURE %%

\section*{Discussion}

In a crystalline host, implanted \elip\ generally stops at a high symmetry site where its electrostatic potential energy is minimized. It is then coupled to its surroundings via magnetic and electrostatic interactions, yielding a phenomenology very similar to conventional NMR, but with the distinction that the location of the nucleus is not known \textit{a priori}. While the site is important for a detailed interpretation, meaningful inferences can often be made without it.

There is one obvious candidate site within the vdW gap in \crs: the $1b$ site shown in \Cref{fig:CrSe2}, which corresponds to the Li site in LiCrSe$_2$ \cite{1971-Laar}. The appearance of two resonances (\Cref{fig:SLR_Res}b) suggesting two distinct sites, is therefore surprising. Since \eli\ is a quadrupolar nucleus, its spin is coupled to the local electric field gradient (\efg) which splits the resonance into a quartet of quadrupolar satellites \cite{2015-MacFarlane}. In a non-cubic crystal, the \efg\ is finite at all sites, but neither resonance in \crs\ appears quadrupole split. However, the \efg\ is probably small at sites in the vdW gap. For example, there is also no resolved quadrupolar splitting in another TMD, 2\textit{H}-NbSe$_2$ \cite{2006-Wang}, where the \eli\ line is much narrower. This is not universal, though, since in some vdW crystals, there is a small splitting \cite{2019-McFadden}, but such splittings are less than the linewidth in \crs. Similarly, conventional $^7$Li NMR in closely related LiCrS$_2$ found no quadrupole splitting, despite comparable quadrupole moments of $^7$Li and $^8$Li \cite{1977-Silbernagel}. We conclude that the \eli\ sites are characterized by a small \efg, suggesting location(s) in the vdW gap.

There is precedent for implanted \elip\ to populate magnetically inequivalent sites in materials. For example, in elemental Au and Ag, it stops in two cubic sites that are distinguished by their magnetic Knight shift \cite{2004-Morris,2008-Parolin}. In \crs, there are two high symmetry sites: one coordinated by six Se atoms in a quasi-octahedron ($1b$) shown in \Cref{fig:CrSe2} and a less stable quasi-tetrahedral ($2d$) site \cite{1971-Laar}. The barrier for \elip\ to migrate from $2d$ to neighboring $1b$ should be small \cite{2000-Van}, so that if the (presumably metastable) $2d$ site were occupied at low temperature, we would expect a site change below 300 K. However, in contrast to Au and Ag, the evolution of the resonance spectrum in \Cref{fig:SLR_Res}b does not show an amplitude trade-off characteristic of a site change.

Instead, the more shifted line broadens and diminishes while the amplitude of the other line does not change appreciably. The temperature independent amplitude of the \slr\ data show that this is not a signal ``wipe-out'' from a divergence of $1/T_1$. It probably reflects a transition to a very broad line that is invisible due to the limited RF amplitude $B_1\sim 0.5$ G. The loss of this line coincides with the range of the two structural/magnetic transitions at 190 and 157 K \cite{2016-Sugiyama}. Superlattice electron diffraction peaks \cite{2014-Kobayashi} below the structural transition indicate a unit cell tripling that results in a multiplicity of similar, but distinct \elip\ sites, which must further contribute to the broadening. Remarkably, both structural transitions are also associated with an \textit{increase} in the susceptibility $\chi_0$ \cite{2014-Kobayashi}. Thus, any distribution in coupling to $\chi_0$ will also cause an amplified magnetic broadening in the lower temperature phase. Transverse field \msr\ also shows a significant broadening on cooling towards $T_{N1}$ \cite{2018-Sugiyama}.

The strongly shifted line is consistent with interstitial \elip\ in the paramagnetic state of \crs. Its shift implies a coupling to the Cr spins polarized by the applied field. In the linear response regime, the shift $\delta_c$ is proportional to the Cr spin susceptibility, $\delta_c = A \chi$ where $A$ is the hyperfine coupling. Using the bulk susceptibility, this relation implies $A \approx 2.1 $ kG/$\mu_{B}$, a relatively small value compared to other metals \cite{2008-Parolin}, again consistent with location(s) in the vdW gap far from the Cr ions. In the magnetically ordered phase, a broad distribution of static magnetic fields is consistent with the fast relaxation of the zero-field \msr\ precession \cite{2016-Sugiyama}. On the other hand, the origin of the less shifted resonance is unclear. We return to this after considering the \slr.

In contrast to the resonances, there is no obvious decomposition of the relaxation into two distinct rates. Instead, there is a broad distribution of rates that is well modeled by the stretched exponential. Three potential contributors to the relaxation are: Cr spin fluctuations, Korringa conduction electron scattering \cite{2009-Hossain}, and diffusive \elip\ hopping. Assuming these act independently, the overall \slr\ rate ($1/T_1$) is given by a simple sum:

%% EQUATION %%

\begin{equation}\label{eq:1/T1}
\frac{1}{T_1} = \left(\frac{1}{T_1}\right)_{\mathrm{Cr}} + \left(\frac{1}{T_1}\right)_{\mathrm{Korr}} + \left(\frac{1}{T_1}\right)_{\mathrm{Diff}}
\end{equation}

%% EQUATION %%

Each term maintains a distinct temperature and field dependence. The Korringa rate increases linearly with temperature, so one might attribute the increase above $\sim$ 200 K in \Cref{fig:T1_Beta}a to this mechanism; however, Korringa relaxation is independent of magnetic field. Instead, $1/T_1$ above 200 K clearly increases with reduced field. The magnitude of the Korringa rate is determined by the square of the hyperfine coupling between the conduction band and the \eli\ nucleus. In TMDs, the band states are concentrated in the trilayers and have little density in the vdW gap, making them highly two dimensional metals. If \elip\ is located in the vdW gap, we expect a weak coupling, and correspondingly slow Korringa relaxation. This is demonstrated in the highly metallic NbSe$_2$, where \eli\ exhibits an extremely slow, but linearly temperature dependent relaxation rate \cite{2006-Wang}. Based on this, we conclude that any effect of $(1/T_1)_{\mathrm{Korr}}$ is overwhelmed by the contributions from $(1/T_1)_{\mathrm{Diff}}$ and $(1/T_1)_{\mathrm{Cr}}$.

The field dependence of $1/T_1$ above 200 K suggests diffusive relaxation. Mobility of Li$^+$ in this temperature range is consistent with room temperature alkali intercalability and ionic conductivity of many TMDs \cite{1978-Whittingham}, including closely related CrS$_2$ \cite{1977-Silbernagel}. Diffusive relaxation in solids is related to the site-to-site hop rate $\nu_{\mathrm{hop}}$. Ideally, one could translate the measured \slr\ rate into a value for $\nu_{\mathrm{hop}}$, but this depends on an unknown coupling constant. When $\nu_{\mathrm{hop}}$ matches the NMR frequency, $1/T_1$ is maximized at a Bloembergen-Purcell-Pound (BPP) peak \cite{1948-Bloembergen}. From this, one can determine the coupling constant and measure $\nu_{\mathrm{hop}}$ as well as the associated activation energy barrier, $E_a$, unambiguously \cite{2019-McFadden,2017-McFadden}. However, absence of a clear $1/T_1$ maximum at high temperatures prevents extracting quantitative information from this BPP model. Nonetheless, the increasing relaxation above 200 K probably stems from the low temperature flank of such a BPP peak not far above room temperature. The field dependence of $1/T_1$ and high temperature resonance narrowing are consistent with this. 

We now consider the evolution of $1/T_1$ below 200 K. Diffusive ionic hopping is an activated process, so that $\nu_{\mathrm{hop}}$ slows continuously with reduced temperature, and one expects $(1/T_1)_{\mathrm{Diff}}$ to fall exponentially below the BPP peak. The opposite trend in $1/T_1$ at low $T$ indicates the predominance of magnetic relaxation due to Cr spin fluctuations below 200 K. This attribution is confirmed by the large peak in the rate near the {\it lower} magnetic transition $T_{N2} \approx 20 $ K \cite{2016-Sugiyama}. Remarkably, there is almost no feature at the upper magnetic transition $T_{N1}$, nor at the structural transition which is clearly evident in $\chi_0(T)$. This is reminiscent of the behavior of the less shifted resonance, which is practically unaffected by the same high temperature transitions, while showing a substantial shift below $T_{N2}$.

We now discuss the absence of a noticeable feature in $1/T_1$ at $T_{N1}$. At highly symmetric sites in an antiferromagnet, the internal field due to the ordered moments may cancel. This is the case for the (dipolar) fields at the $1b$ site in the vdW gap for some of the candidate magnetic structures considered to interpret the \msr\ data \cite{2016-Sugiyama}. In an incommensurate SDW state, however, there is no such site, and the absence of a feature at the transition suggests the ordered SDW moment is small and/or the coupling to the \eli\ is particularly weak. This is in sharp contrast to  the zero field \msr, which shows clear evidence of static magnetic order below $T_{N1}$. This difference  suggests that either the lattice sites for $\mu^+$ and \elip\ are quite different, or that there is a significant influence of the applied field. Note that the muon adopts several distinct environments in the magnetic phase, giving rise to multiple precession frequencies. This is quite different than the two resonances we find in the paramagnetic phase at \textit{high temperature}.

Although there are unexplained aspects of the low temperature relaxation, it is clearly magnetic. Though it decreases with increasing temperature, it remains important in the paramagnetic phase. We now use this understanding to develop a simple model of $1/T_1$ at high temperatures. Above 200 K, the magnetic relaxation $(1/T_1)_{\mathrm{Cr}}$ will be approximately independent of temperature, similar to the paramagnetic state of LiCrO$_2$ \cite{Alexander2007}. We then ascribe the high temperature increase to $(1/T_1)_{\mathrm{Diff}}$, which we model with a simple Arrhenius law:

\begin{equation}\label{eq:1/T1arr}
\frac{1}{T_1} = C + A \mathrm{e}^{-E_{a}/kT} 
\end{equation}

Here, the ionic hopping is parameterized by the pre-exponential factor $A$, the activation barrier $E_a$, and Boltzmann constant $k$. An estimate for the magnetic relaxation rate $C$ is obtained from the minimum observed $1/T_1$. In the slow fluctuation limit, where the ionic hop rate is much less than the NMR frequency, one expects a stronger field dependence in the case of purely diffusive relaxation \cite{1948-Bloembergen}, but the presence of the magnetic term $C$ limits the variation with field. A global fit at \textit{both} fields (parameters $A$ and $C$ free) to this model provides the curves in \Cref{fig:T1_Beta}a and a shared $E_a = 0.12(1)$ eV, in the range expected from other vdW materials \cite{2019-McFadden}.

Finally, we return to the two-component resonance spectrum. It is challenging to reconcile the presence of two \eli\ sites that are differently sensitive to the two magnetic transitions. The unshifted line does not appear to be a background (e.g., from \eli\ stopping in the grease between the crystals) because its amplitude is too large and it would also not be sensitive to $T_{N2}$. The growth of the shifted line may represent a site change for \eli. However, there is no accompanying amplitude reduction for the unshifted line. The coincidence of a site change with the structural transitions also appears fortuitous. Rather, we speculate that the unshifted line may correspond to \eli\ stopping in a magnetically distinct near-surface layer, similar to a previous observation in metallic palladium, where it may be due to surface adsorbates \cite{MacFarlane2012}. Similar features have also been found in conventional NMR of Pt nanoparticles coated with chemisorbed carbon monoxide \cite{1982-Rhodes}.

It is also difficult to accurately estimate the amount of implanted \eli\ corresponding to the two resonances. However, the fractions must be comparable, as both resonance amplitudes are large. With an average implantation depth of nearly 100 nm, a negligible fraction of \eli\ are in the topmost atomic layers. To account for the amplitude of the unshifted line, one would require a surface layer extending several tens of nanometers beneath the surface. It remains to be understood how \eli\ in a distinct surface layer would be proximally sensitive only to the lower transition. This might be due to a ferromagnetic component to the low $T$ order which would extend its influence further beyond the magnetically ordered volume. In any case, this suggestion could be tested directly by \textit{reducing the implantation energy}, which should increase the relative amplitude of the unshifted line.

Collectively, these results highlight the unusual magnetic properties of \crs. Although the presence of \lip\ diffusion above 200 K (in the paramagnetic state) is unsurprising, the lack of any variation in the SLR rate near a previously reported \cite{2016-Sugiyama} magnetic transition ($T_{N1}$) is indeed unusual among antiferromagnetic materials, as both conventional NMR and \bnmr\ are typically sensitive to such a feature.

\section*{Conclusions}
   
We have demonstrated the application of implanted ion \elip\ \bnmr\ for an anomalous TMD, \crs. Though strongly magnetic, we can follow the resonances and \slr\ over the entire temperature range between 4 and 300 K. From the resonances, the lack of quadrupole splitting and the weak hyperfine coupling to the paramagnetic response of the Cr ions suggests \elip\ implantation sites within the vdW gap. The broad lines and stretched exponential relaxation reflect considerable inhomogeneity, but the presence of two environments indicated by the two resolved resonances at high temperatures appears to require an additional source of inhomogeneity. The \slr\ rate $1/T_1$ increases monotonically above 200 K, which is likely due to diffusion of the probe ion. It also passes through a peak near the low temperature magnetic transition. In contrast, a broad minimum is noted with surprisingly little change at the structural and magnetic transitions in the interval 150 to 200 K.

\section*{Conflicts of Interest}

There are no conflicts to declare.   

\section*{Acknowledgments}

We are grateful for outstanding technical and experimental assistance from R. Abasalti, D. Arseneau, S. Daviel, and D. Vyas. We thank L. Schoop for useful discussions. J. Sugiyama was supported by Japan Society for the Promotion Science (JSPS) KAKENHI Grants No. JP26286084 and JP18H01863. S. Kobayahshi and K. Yoshimura were supported by JSPS KAKENHI Grant No. JP18KK0150. QuEST fellowship support was provided to D. Fujimoto, V. L. Karner, and J. O. Ticknor. IsoSiM fellowship support was extended to A. Chatzichristos and R. M. L. McFadden. R. F. Kiefl and W. A. MacFarlane acknowledge NSERC funding.

\bibliography{CrSe2-references} 

\end{document}